# Improving diagnosis and prognosis of lung cancer using vision transformers: A scoping review

Hazrat Ali[1*], Farida Mohsen[1], Zubair Shah[1*]
College of Science and Engineering, Hamad Bin Khalifa University, Qatar Foundation, Doha, Qatar.
Email: haali2@hbku.edu.qa, famo39885@hbku.edu.qa, zshah@hbku.edu.qa

**Abstract**

**Background:**
Vision transformer-based methods are advancing the field of medical artificial intelligence and cancer imaging, including lung cancer applications. Recently, many researchers have developed vision transformer-based AI methods for lung cancer diagnosis and prognosis.

**Objective:**
This scoping review aims to identify the recent developments on vision transformer-based AI methods for lung cancer imaging applications. It provides key insights into how vision transformers complemented the performance of AI and deep learning methods for lung cancer. Furthermore, the review also identifies the datasets that contributed to advancing the field.

**Methods:**
In this review, we searched Pubmed, Scopus, IEEEXplore, and Google Scholar online databases. The search terms included intervention terms (vision transformers) and the task (i.e., lung cancer, adenocarcinoma, etc.). Two reviewers independently screened the title and abstract to select relevant studies and performed the data extraction. A third reviewer was consulted to validate the inclusion and exclusion. Finally, the narrative approach was used to synthesize the data.

**Results:**
Of the 314 retrieved studies, this review included 34 studies published from 2020 to 2022. The most commonly addressed task in these studies was the classification of lung cancer types, such as lung squamous cell carcinoma versus lung adenocarcinoma, and identifying benign versus malignant pulmonary nodules. Other applications included survival prediction of lung cancer patients and segmentation of lungs. The studies lacked clear strategies for clinical transformation. SWIN transformer was a popular choice of the researchers; however, many other architectures were also reported where vision transformer was combined with convolutional neural networks or UNet model. Researchers have used the publicly available lung cancer datasets of the lung imaging database consortium and the cancer genome atlas. One study used a cluster of 48 GPUs, while other studies used one, two, or four GPUs.

**Conclusion:**
It can be concluded that vision transformer-based models are increasingly in popularity for developing AI methods for lung cancer applications. However, their computational complexity and clinical relevance are important factors to be considered for future research work. This review provides valuable insights for researchers in the field of AI and healthcare to advance the state-of-the-art in lung cancer diagnosis and prognosis. We provide an interactive dashboard on [lung-cancer.onrender.com/](lung-cancer.onrender.com/).

# 1   Introduction

Lung cancer is a highly prevalent and fatal form of cancer globally [1], [2]. Over the last few decades, medical imaging techniques have played an increasingly vital role in diagnosing, prognosis, survival prediction, and early detection of lung cancer, eventually aiding in effective cure and prevention. Such techniques make use of lung computed tomography (CT), X-rays, positron emission tomography (PET), and magnetic resonance imaging (MRI). Traditionally, medical images in clinical work have been interpreted and analyzed by trained radiologists who use their expertise and experience to make accurate diagnoses. However, the manual interpretation of medical images can be time-consuming, prone to human error, and affected by intra-observer as well as inter-observer variability.

Artificial intelligence (AI) methods, particularly deep learning models, have played a vital role in automating image processing in the past few years and have been gaining increasing attention in medical imaging [3], [4]. AI methods dominated by convolutional neural networks (CNNs) [5], [6] have revolutionized the realm of medical imaging with their capability of learning complex representations enabling the automated diagnosis of diseases and the detection of abnormalities. They have demonstrated remarkable improvements in various medical imaging applications and modalities, including MRI [7], [8], CT [9], endoscopy [10], and radiography [11], [12], to name a few. However, the advent of transformers apprised researchers of CNNs' major drawback, i.e., the inability to capture long-range dependencies such as the extraction of contextual information and the non-local correlation of objects.

Recently, Dosovitskiy et al. [13] sought to apply the success of transformers in natural language processing to image processing. They developed a vision transformer to capture long-term dependencies within an image by treating image classification as a sequence prediction task for a series of image patches. On several benchmark datasets, the vision transformer and its derived instances demonstrated state-of-the-art (SOTA) performance and gained popularity in several computer vision tasks, including classification [13], segmentation [14], and detection [15]. The use of vision transformers has also been cross-pollinated into the medical image field, where they are used for image segmentation [16], synthesis [17], and disease diagnosis, resulting in SOTA performances. For lung cancer imaging applications, the use of vision transformers has gained attention for different applications, including cancer classification, tumor segmentation, nodule detection, and survival prediction. Much new vision transformer-based AI methods for lung cancer imaging applications have recently been published by researchers.

Our scoping review aims to present a comprehensive overview of the recent studies that developed vision transformer-based AI methods for lung cancer imaging. While there are a few related reviews in the literature [18]–[21]; they differ in their focus and coverage. For example, the review in [18] covers the applications of vision transformers in medical imaging; however, it is not specific to lung cancer imaging applications and covers many different medical imaging applications. Similarly, the reviews in [19], [20] cover other AI methods for cancer imaging but do not include vision transformers, while the review in [21] covers AI methods for lung cancer applications but covers only pathology imaging and does not cover all the imaging modalities. Besides, it does not cover the recent developments of vision transformers for lung cancer imaging, as the review was published much earlier. To the best of our knowledge, no review study focuses specifically on the utilization of vision transformers for medical imaging in lung cancer. Therefore, our review is the first comprehensive review that focuses specifically on the use of vision transformers for medical imaging in lung cancer, providing a thorough overview of the current state of the field. Table 1 shows the literature review comparison.

Table 1. Literature comparison with previous review studies

| Reference | Year | Scope and coverage | Differences with our review |
|---|---|---|---|
| Transformers in Medical Image Analysis [18] | August 2022 | It focuses on the use of transformers for various medical imaging applications. It is not specific to lung cancer imaging. It does not cover many recent studies on vision transformers in lung cancer imaging. | Our review focuses specifically on the use of vision transformers for lung cancer imaging. Our review covers many recent studies published in later 2022. |
| Artificial intelligence in lung cancer: current applications and perspectives [19] | November 2022 | It focuses on the current state of AI in lung cancer. It covers traditional machine learning and deep learning methods for lung nodule detection and segmentation. It does not cover vision transformer-based approaches. | Our review focuses specifically on the use of vision transformers for lung cancer imaging. |
| Artificial intelligence techniques for cancer detection in medical image processing: A review [20] | May 2021 | It focuses on a broad range of AI methods. It covers many different types of cancer imaging applications. It is not specific to vision transformers. It is not specific to lung cancer. It does not cover many recent studies. | Our review focuses specifically on the use of vision transformers for medical imaging in lung cancer. Our review is specific to vision transformers. Our review is specific to lung cancer. Our review covers recent studies. |
| Artificial Intelligence in Lung Cancer Pathology Image Analysis [21] | November 2019 | It focuses on AI methods for pathology image analysis of lung cancer. It does not cover vision transformers. It does not cover recent studies. | Our review covers the use of transformers for medical imaging in lung cancer. Our review covers different imaging modalities, including pathology and CT scans. Our review is specific to vision transformers. Our review covers recent studies. |
| Recent advances of Transformers in medical image analysis: A comprehensive review [60] | March 2023 | It focuses on transformers for various medical imaging applications. It is not specific to lung cancer imaging and covers many studies on COVID-19. It does not cover many recent studies on vision transformers in lung cancer imaging. | Our review focuses specifically on the use of vision transformers for lung cancer imaging. Our review covers many recent studies published in later 2022. |
| Machine Learning for Lung Cancer Diagnosis, Treatment, and Prognosis [61] | October 2022 | It covers different machine learning and deep learning techniques. It does not cover any study on vision transformer applications. It is not specific to imaging modality and covers studies on different data modalities for lung cancer. | Our review focuses on vision transformer applications in lung cancer imaging. Our review focuses on imaging data for lung cancer. |

The primary aim of our scoping review is to synthesize scientific literature by answering the following research questions, as listed in Figure 1C.

We are confident that this review will provide a comprehensive text on the recent developments in vision transformer-based lung cancer imaging applications.

# 2 Methods

In the review, we followed the PRISMA-ScR (Preferred Reporting Items for Systematic Reviews and Meta-Analyses) [22] guidelines to perform the study search and synthesis of the data.

## 2.1 Search Strategy

An extensive search of scientific databases, including PubMed, Scopus, IEEE Xplore, Google Scholar, and MEDLINE (via PubMed), was conducted to identify the relevant studies. The study search was performed on December 21, 2022. Reference list checking was also performed for additional relevant studies. Only the first 150 relevant studies from Google Scholar were considered for the review, as search results beyond this number rapidly lost relevance and were not pertinent to the scoping review topic. The search terms were defined through consultation with domain experts and on the basis of the previous literature. The search terms included terms based on the target anatomy (e.g., lung cancer) and the intervention (e.g., transformers). The detailed search strings used in the study can be found in Appendix 1.

## 2.2 Search eligibility criteria

In this scoping review, we focused on exploring the recent advancements and applications of vision transformers in lung cancer medical imaging. We analyzed studies published until December 2022 in English that involved the utilization of vision transformers for various purposes related to lung cancer imaging, such as classification of lung cancer types, prediction of the cancer growth, detection of nodule, survival prediction of lung cancer patients, and segmentation of lungs. Studies that used any medical imaging modality such as MRI, CT, X-ray, and histopathology images were considered. Only original research published in peer-reviewed journals, conference proceedings, or book chapters was considered.

Studies that did not use vision transformers specifically but utilized other deep learning methods, such as CNNs and Generative Adversarial Networks (GANs), were excluded from the review. Additionally, studies that used transformers for non-imaging data, such as text data and electronic health records (EHRs), were excluded. Moreover, studies that used transformers for cancers other than lung cancer were also excluded. Studies identified as non-English text, review articles, preprints, editorials, proposals, conference abstracts, commentaries, and editor letters were also excluded. No restrictions were in place on the country of publications, the models' complexity, the reported methods' performance, and the modality of imaging data.

## 2.3 Study selection and data extraction

We used the Rayyan web-based review management tool [23] to conduct the initial screening and study selection process. One reviewer (H.A.) performed the literature search. After eliminating duplicates, two reviewers (F.M.) and (H.A.) independently screened the titles and abstracts of the studies to identify eligible studies. The studies that successfully passed the initial title and abstract screening were selected for the full-text screening phase. Any disagreements during the process were resolved through discussion and through validation by a third reviewer (Z.S.). An evidence form was created and tested on three studies to establish a systematic and precise data extraction process (also see Appendix 2). Data extracted from the studies included the titles, first author's name, publication date and venue, the country of the first author's institution, the study application, the imaging type, the transformer type, the data source (public or private), data size, the validation methods, and the evaluation metrics. Additionally, information regarding the required hardware resources was also extracted. Moreover, the studies' challenges and suggested solutions were extracted, along with the challenges encountered and proposed solutions in the studies. Two reviewers (F.M. and H.Z.) conducted the data extraction, and any discrepancies were resolved through discussions and mutual consensus.

## 2.4 Data Synthesis

We followed a narrative approach to synthesize the data extracted from the included studies. We categorized the data in terms of the specific tasks addressed in them, such as classification of lung cancer type, prediction of the course of cancer, survival prediction of the cancer patients, and segmentation of lungs. Based on the models developed in the included studies, we categorized them into those using 3D models and those using 2D models. We also cataloged the studies based on the use of public versus privately developed datasets, the method of validation of the results, and the reproducibility of the results.

# 3 Results
## 3.1 Search Results

The search retrieved 314 studies. However, 92 studies were duplicates that we removed. We removed 183 studies according to the exclusion/inclusion criteria in the title and abstract screening phase. In the remaining 39 studies, we removed eight more studies after the full-text reading phase, as they did not fulfill the inclusion criteria. We added three additional studies through forward/backward referencing. Finally, we included 34 unique studies [24]–[57] in the review (also see Appendix 3 for all the included studies). Figure 1 shows the flowchart for the different phases of the study selection and the number of studies retained in each phase. Readers may access an interactive dashboard on [lung-cancer.onrender.com/](lung-cancer.onrender.com/). (Loading may take up to 60 seconds)

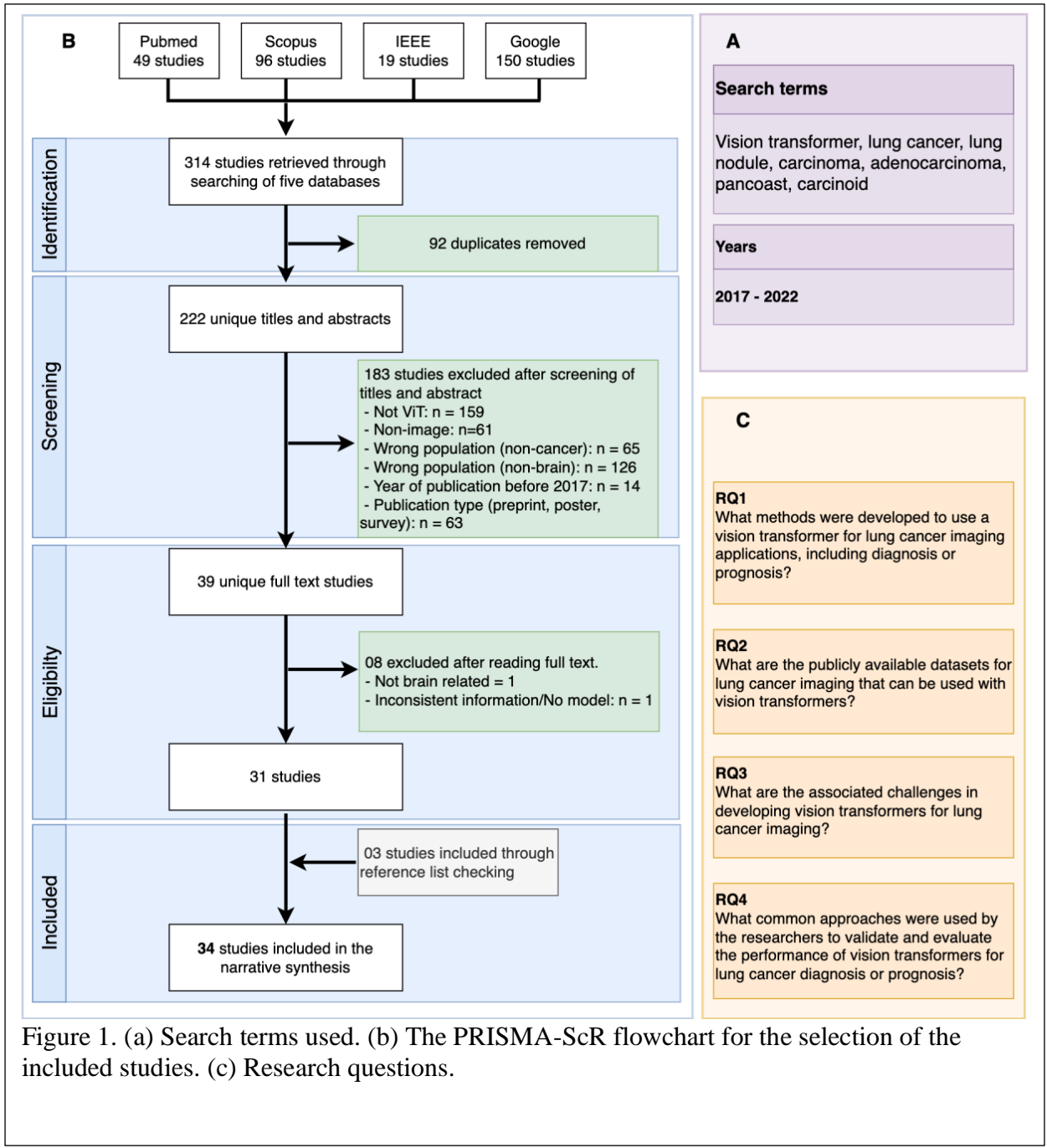

Figure 1. (a) Search terms used. (b) The PRISMA-ScR flowchart for the selection of the included studies. (c) Research questions.

### 3.2 Demographics of the included studies

In the included studies, half of the studies (n=17) were journal articles, while 16 studies were published in conference proceedings. Only one study was published as a thesis. Most of the studies (n=28) were published in 2022, four studies were published in 2021, and only two studies were published in 2020. Of the articles published in 2022, five studies were published in June, five were published in September, and four were published in August. Of the four studies published in 2021, no study was published in the first eight months. The included studies were published by researchers from seven different countries (first-author country affiliation). Researchers from China published almost two-third (n=21) of the studies, while researchers from the USA published approximately one-fourth (n=8) of the studies. Researchers from India, Saudi Arabia, Pakistan, Canada, and South Korea published one study each. Figure 2 shows a summary of the year-wise, and Figure 3 shows the country-wise demographics of the included studies. Table 2 summarizes the demographics of the included studies.

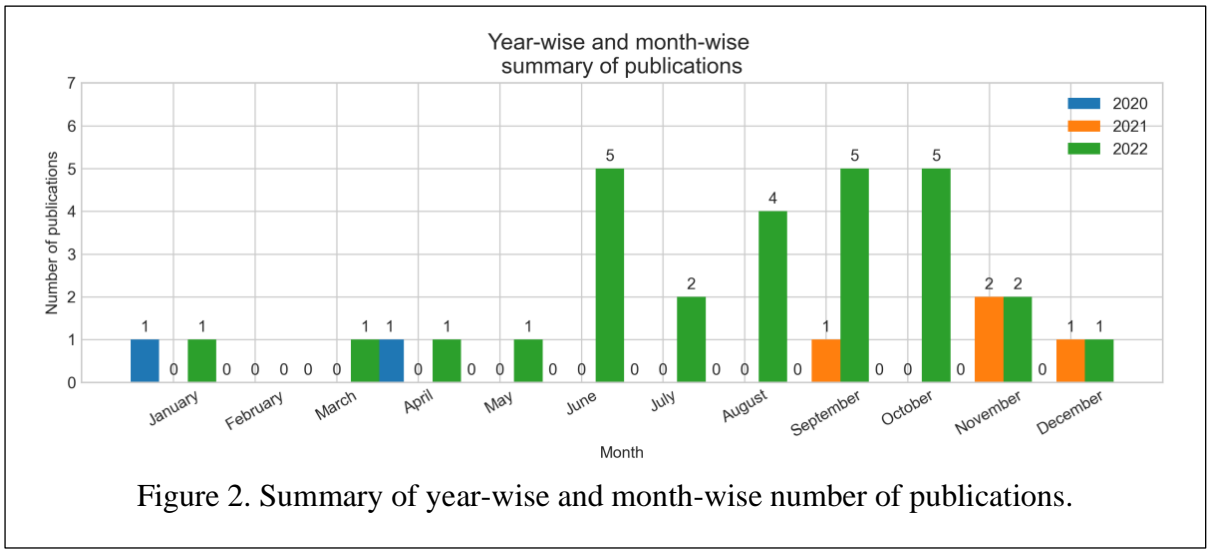

Figure 2. Summary of year-wise and month-wise number of publications.

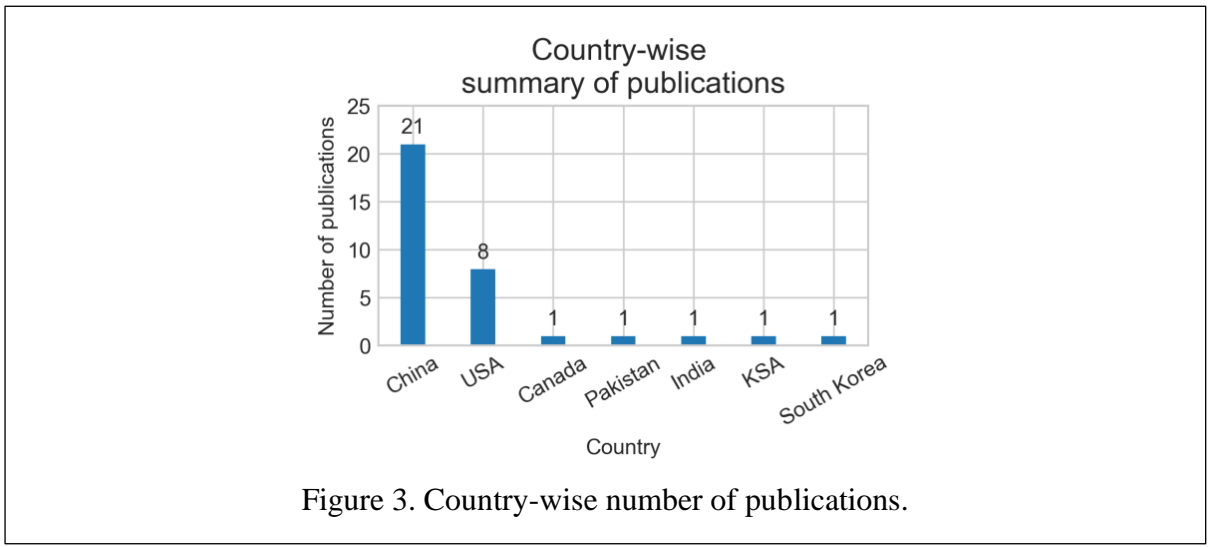

Figure 3. Country-wise number of publications.

Table 2. Demographics of the inlcuded studies.

| Year | Year | Month | Number of studies |
|---|---|---|---|
| | 2022 (n=28) | January | 1 |
| | | March | 1 |
| | | April | 1 |
| | | May | 1 |
| | | June | 5 |
| | | July | 2 |
| | | August | 4 |
| | | September | 5 |
| | | October | 5 |
| | | November | 2 |
| | | December | 1 |
| | 2021 (n=4) | September | 1 |
| | | October | 2 |
| | | December | 1 |
| | 2020 (n=2) | January | 1 |
| | | April | 1 |
| Countries | Country | | Number of studies |
| | China | | 21 |
| | USA | | 8 |
| | Canada | | 1 |
| | India | | 1 |
| | Saudi Arabia | | 1 |
| | South Korea | | 1 |
| | Pakistan | | 1 |
| Type of publication | Venue | | Number of studies |
| | Journal | | 17 |
| | Conference | | 16 |
| | Thesis | | 1 |

### 3.3 Main tasks addressed in the studies

In the 34 studies included in this review, one-third of studies (n=11) [24]–[35] performed classification of different types of lung cancers. Nearly half of the studies (n=15) [35], [43], [45]–[57] used vision transformer-based models to predict the growth of tumors or the course of cancer. Of these, eight studies [35], [43], [48], [53]–[57] developed ViT-based models for survival prediction of lung cancer patients. Six studies [36]–[43] addressed the task of segmentation of tumor or lung nodules. One study [44] performed lung nodule detection. Few studies performed more than one task. For example, one study [35] performed the classification of lung cancer types and reported performance for survival prediction too. Similarly, one study [42] performed segmentation of lungs and detection of nodules, and one study [43] reported segmentation of lungs and survival prediction of patients. Figure 4 shows a mapping of the different tasks addressed in the included studies.

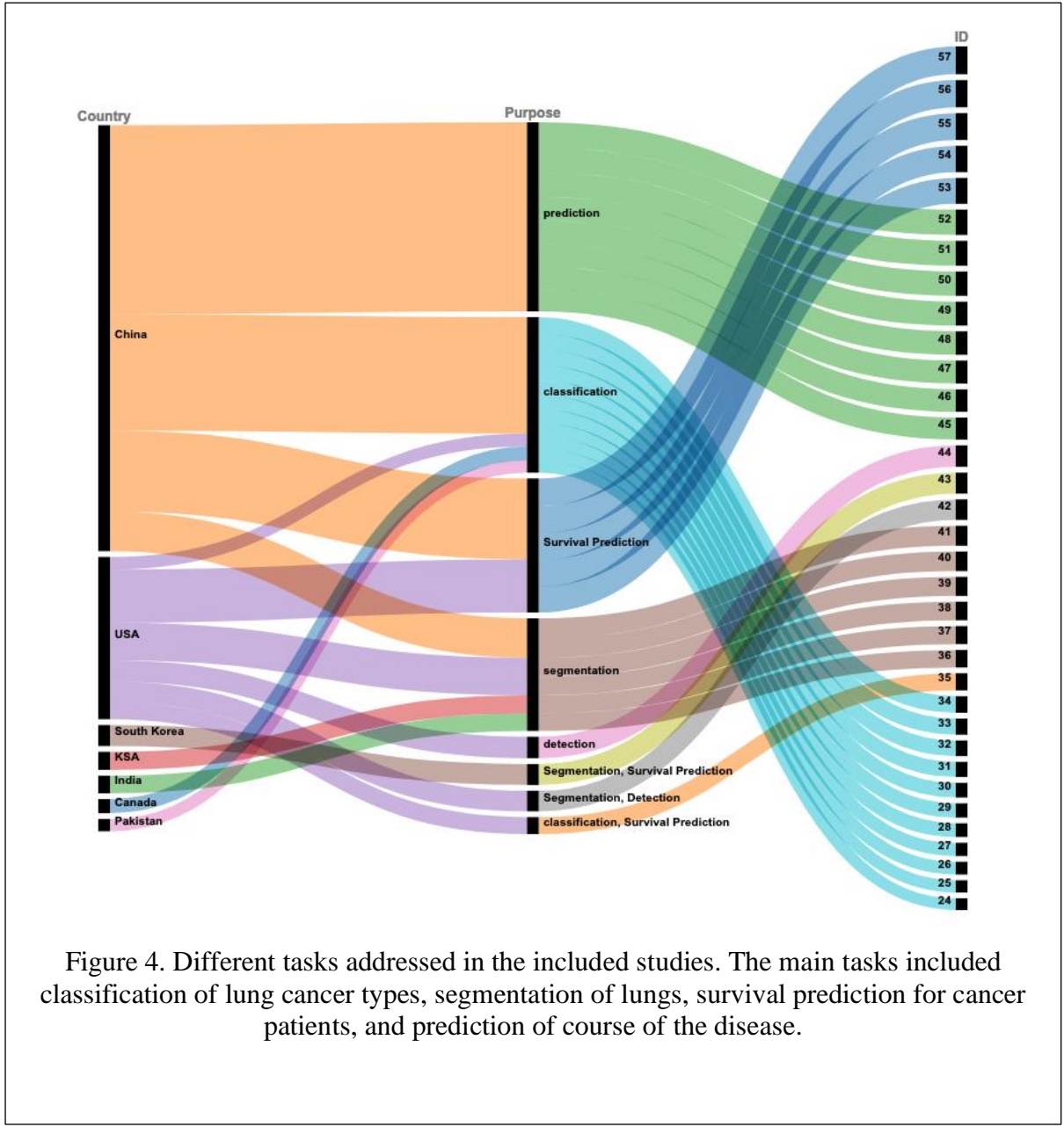

Figure 4. Different tasks addressed in the included studies. The main tasks included classification of lung cancer types, segmentation of lungs, survival prediction for cancer patients, and prediction of course of the disease.

## 3.4 Key implementation details

In the included studies, vision transformers were combined with CNNs, UNet, or graph networks. In the included studies, seven studies [25], [28], [34], [37], [51], [57] combined vision transformers with CNNs, three studies [36], [38], [52] used vision transformers in combination with UNet model, one study [41] combined both CNN and UNet with vision transformer, one study combined vision transformer with ResNet model. One study [24] combined the mask R-CNN model with a vision transformer to perform segmentation followed by classification. Two studies [48], [55] explored the use of graph networks in combination with vision transformers. Six studies [24], [28], [32], [46], [49], [50] used SWIN transformer as their backbone transformer architecture.

Almost half of the studies (n=18) [24], [26]–[28], [30], [31], [33], [35]–[37], [41], [42], [45]–[49], [53] reported that their implementation was in Pytorch framework, while one study [25] reported the use of TensorFlow and Keras frameworks. The remaining studies did not specify the framework used.

Three studies [24], [26], [40] reported the use of a single Nvidia RTX 2080Ti GPU that usually comes with 11 GB memory, while one study [53] reported the use of four Nvidia RTX 2080Ti GPUs with 12 GB memory. Two studies [25], [36] reported the use of Nvidia P100 GPU, where the authors in [25] accessed the GPU via the Kaggle computational platform. Four studies [28], [39], [42], [57] reported the use of Nvidia V100 GPUs. Of these, one study [39] used four GPUs, one study [42] used two GPUs, and one study [57] used a single V100 GPU. Three studies [27], [33], [37] used a single Nvidia RTX3090 GPU, while one study [41] used a combination of two Nvidia RT3090 GPUs. Three studies [45], [54], [55] used a single Nvidia GTX 1080 or 1080Ti GPU with 11 GB memory. One study [47] used Nvidia Titan-XP GPU. The largest number of GPUs usage was reported by [44], who used 48 Nvidia V100 GPUs. The remaining studies did not provide information on GPU usage.

### 3.5   Types of data used in the studies

In the included studies, 22 studies reported the publicly available use of data, six studies reported experiments on privately collected data, and six studies used both public and private datasets. In the included studies, 23 developed models for 2D image data while 11 developed models for volumetric data. Nearly two-third (n=21) of the included studies used computed tomography (CT) scans of lung, while one-third (n=11) studies used histopathology or whole slide images of lungs. One study used PET, while another used CT and MRI scans. Table 3 summarizes the use of types of data in the included studies. Figure 5 shows the number of studies that used different modalities of data. Figure 6 shows the Venn diagram for the number of studies using public versus private data.

Table 3. Summary of types of data used.

| | | Used by |
|---|---|---|
| Availability of dataset | Public data (n=22) | [25]–[28], [30]–[32], [34]–[40], [42], [44], [51]–[56] |
| | Private data (n=6) | [24], [29], [33], [41], [46], [49] |
| | Public and private data (n=6) | [43], [45], [47], [48], [50], [57] |
| Dimensionality of data | 2D models (2D data) (n=23) | [24], [26]–[28], [30]–[32], [34]–[37], [40], [41], [47]–[50], [52]–[57] |
| | 3D models (volumetric data) (n=11) | [25], [29], [33], [38], [39], [42]–[46], [51] |
| Modality of image data | CT (n=21) | [29], [30], [32], [33], [33], [34], [36]–[42], [44]–[52] |
| | Histopathology (n=11) | (n=11) [24]–[28], [35], [53]–[57] |
| | PET (n=1) | [43] |
| | CT and MRI (n=1) | [39] |

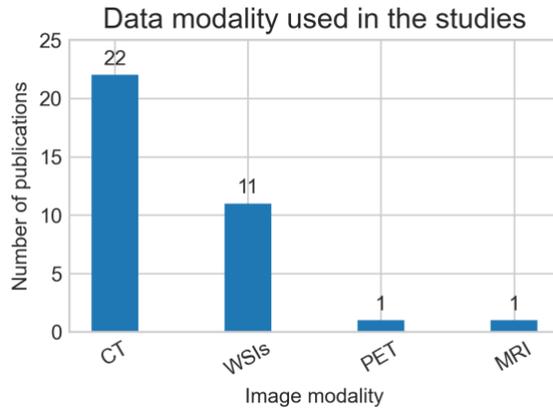

Figure 5. Different data modalities used in the included studies.

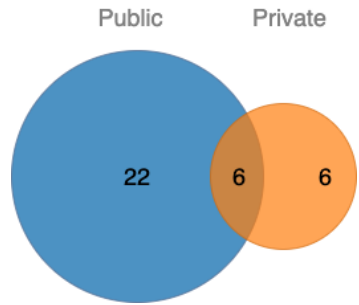

Figure 6. Venn diagrams showing the contribution of public versus private datasets used

### 3.6 Datasets used in the studies

In the included studies, six studies [30]–[32], [40], [44], [52] the Lung Imaging Database Consortium (LIDC-IDRI) dataset, five studies [27], [28], [54]–[56] used The Cancer Genome Atlas (TCGA) datasets, four studies [34], [36]–[38] used the LUNA16 dataset. Table 4 summarizes the datasets used in the included studies along with the URL for the publicly available datasets.

Table 4. Datasets used in the included studies

| Dataset name | URL | Used by |
|---|---|---|
| LC25000 | https://github.com/tampapath/lung_colon_image_set | [24] |
| National Lung Screening Trial (NLST) | https://www.cancer.gov/types/lung/research/nlst | [26], [47], [50] |
| Non-small cell lung cancer (NSCLC) | https://www.cancer.gov/about-nci/organization/ccg/research/structural-genomics/tcga/studied-cancers/lung-adenocarcinoma | [27], [48] |
| Lung Squamous Cell Carcinoma (TCGA-LUSC) | https://wiki.cancerimagingarchive.net/pages/viewpage.action?pageId=16056484 | [44], [54], [55] |
| Lung Adenocarcinoma (TCGA-LUAD) | https://www.cancerimagingarchive.net/collections/tcga-luad/ | [28], [53], [56] |
| Lung Imaging Database Consortium (LIDC-IDRI) | https://wiki.cancerimagingarchive.net/pages/viewpage.action?pageId=1966254 | [30]–[32], [40], [44], [52] |
| LUNA16 | https://luna16.grand-challenge.org/Data/ | [34], [36], [37], [44] |
| Tianchi Lung Nodule Detection dataset | https://tianchi.aliyun.com/competition/entrance/231601/introduction | [34] |
| Cbioportal | https://www.cbioportal.org/ | [35] |
| Medical Segmentation Decathlon | http://medicaldecathlon.com/ | [42] |
| LUNG1 | https://wiki.cancerimagingarchive.net/display/Public/NSCLC-Radiomics | [43] |
| LUNGx | https://wiki.cancerimagingarchive.net/display/Public/SPIE-AAPM+Lung+CT+Challenge | [44] |
| NSCLC Radiogenomics | https://wiki.cancerimagingarchive.net/display/Public/NSCLC+Radiogenomics | [27], [39], [48] |
| NLSTt (derived from NLST) | https://github.com/liaw05/STMixer | [51] |
| Shanghai pulmonary hospital | Private | [33], [48] |
| Huadong Hospital dataset | Private | [45] |
| West China Hospital of Sichuan University | Private | [46] |
| Shanxi Provincial People's Hospital | Private | [47], [50] |
| CHCAMS | Private | [57] |

### 3.7 Evaluation metrics

The most commonly used evaluation metrics in the included studies were accuracy and area under the ROC curve (AUC), each reported in 16 studies. Other popular metrics were specificity reported in 11 studies, sensitivity reported in nine studies, dice similarity coefficient reported in seven studies, and concordance index reported in six studies. Both precision and recall measures were reported in five studies each. Other metrics were the F1 score, mean absolute error, mean absolute error, root mean square error, and Kappa score, each reported by one study. Figure 7 summarizes the number of studies using different evaluation metrics.

Almost one-third of studies (n=11) reported splitting the data into training, validation, and test sets, while five reported splitting the data into training and test sets only. Similarly, eight studies used a 5-fold cross-validation scheme, while six used a 10-fold cross-validation scheme to evaluate the performance of their methods.

In the included studies, only nine studies [26]–[28], [35], [37] provided a GitHub link for the implementation code.

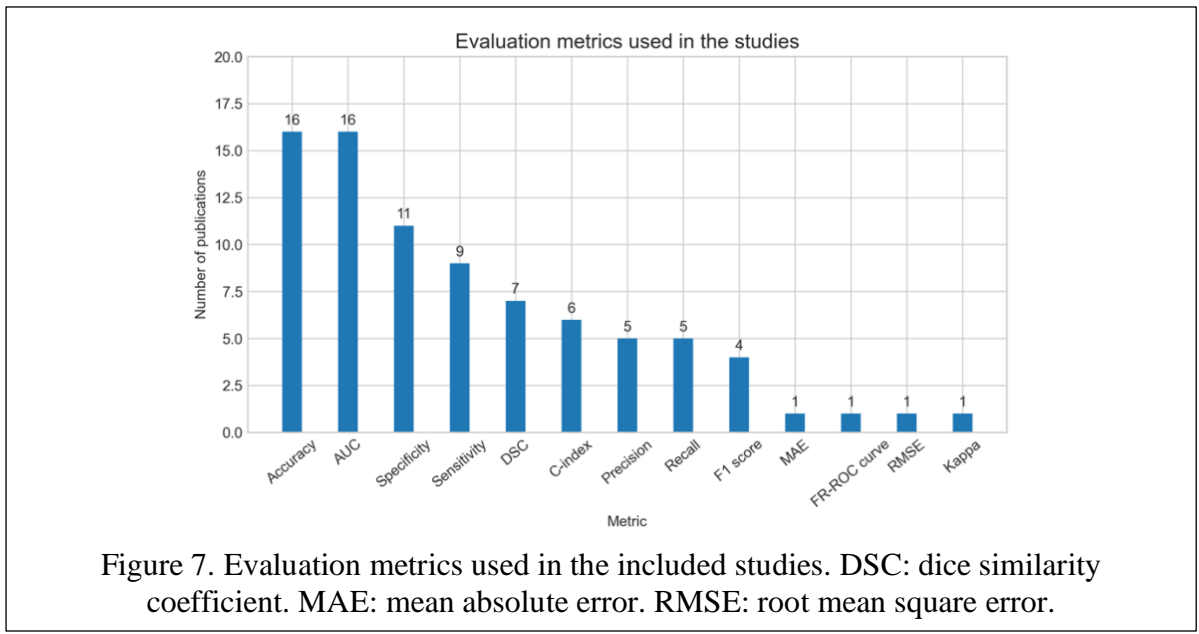

Figure 7. Evaluation metrics used in the included studies. DSC: dice similarity coefficient. MAE: mean absolute error. RMSE: root mean square error.

# 4 Discussion
## 4.1 Principle results

This study provides an overview of recent literature on the utilization of vision transformer-based artificial intelligence models for enhancing the diagnosis, prognosis, and classification of lung cancer. In the review, we did not find any studies before 2020. This is not surprising as vision transformers were proposed in 2017, and their use in medical imaging has recently gained popularity. Most of these studies were published in 2022, reflecting the growing interest in developing vision transformer-based approaches for lung cancer applications. However, the diversity of authors was limited, as researchers from China or the United States authored 85% of the studies.

The popularity of vision transformers for classification tasks has driven a majority of the studies reviewed in this work to employ them for classifying different types of lung cancers. The classification tasks included separating lung squamous cell carcinoma from lung adenocarcinoma, identifying benign versus malignant pulmonary nodules, and determining the invasiveness of lung adenocarcinomas. The studies also used vision transformers to predict lung cancer's severity or growth, thus aiding in survival predictions for patients. Some of the studies were limited to segmenting lung nodules.

Vision transformers effectively capture the long-range context in the input data, while CNNs tend to excel in capturing short-range dependencies. This is why many studies reviewed in this work combined vision transformers and CNNs, either through cascade or parallel connections or by incorporating vision transformer attention mechanisms into CNNs. Researchers integrated vision transformers with UNet or Mask R-CNN models for segmentation tasks. Due to its inherent benefits, the SWIN transformer was frequently used as the backbone architecture for lung cancer imaging applications. The most popular framework for implementing vision transformer-based models in the studies was Pytorch, with all but one study (that used Tensorflow and Keras) reporting the use of Pytorch as the implementation framework.

Vision transformer-based models, in general, are computationally demanding. The computational demands of vision transformer models were addressed in the studies, with some utilizing multiple GPUs and one using a cluster of 48 GPUs, while others demonstrated that implementation on a single GPU was feasible.

Since the most commonly addressed task was the classification of lung cancers; hence, the studies reported accuracy and area under the ROC curve. The concordance index was a common evaluation metric for studies that reported survival prediction. In machine learning models, it is common to split the data in training, validation, and test sets (or training and test sets); however, some studies did not specify the evaluation mechanism and data split.

## 4.2 Practical and Research Implications:

In developing AI models based on vision transformers, the availability of public datasets plays a crucial role. More than two-thirds of the studies utilized publicly available datasets for lung cancer imaging analysis. To encourage further growth in this field, it is imperative to have a rich resource of large-scale public datasets for lung cancer. In our review, the most commonly used imaging modality for lung cancer analysis was CT scans, followed by histopathology images. The use of PET and MRI was found to be less common. The Lung Imaging Database Consortium (LIDC) and The Cancer Genome Atlas (TCGA) offer extensive datasets for lung cancer (and other cancers) for researchers to utilize.

Despite the promising outcomes of vision transformer-based AI methods in analyzing lung cancer, they have limitations, such as the reliance on a significant amount of computational resources, including clusters of numerous GPUs, which may not be accessible to many research laboratories. Moreover, their practical implementation in a clinical setting remains unverified. Hence, there is a pressing need to advance toward developing computationally efficient training methods for vision transformers. The analyzed studies failed to furnish a comprehensive understanding of the interpretability of vision transformer-based models. This information is critical in applications such as predicting the survival of lung cancer patients, as it provides a deeper insight into the progression of the disease and the related risk factors. Additionally, the majority of studies (73%) did not provide access to their implementation code, hindering the ability of other researchers to reproduce the results or build upon the vision transformer-based models for lung cancer analysis. The absence of such links further reduces the reproducibility of the reported studies.

In our review, studies from China and USA dominated the literature where the healthcare tools are advanced, and thus, the new techniques can, in general, be integrated with less effort. However, there is a lack of studies from developing countries. However, there is a scarcity of studies from developing countries. It is anticipated that increased contributions from these countries would aid in addressing the challenges of lung cancer in underdeveloped economies, where the disease is more prevalent due to socio-economic reasons.

The included studies greatly varied in how they reported the different datasets' usage or the number of images for training, validation, or test sets. For example, many studies reported the values of accuracy, sensitivity, specificity, or AUC, the number of samples in the test set varied between them, or the cross-validation strategy differed (or was even absent) in some of the studies. Accordingly, this review does not provide a quantitative summary of the results reported in the included studies for two reasons. First, the review aimed to identify the recent AI methods that used vision transformers for lung cancer imaging applications. Second, the review included many studies that vary in how they report quantitative metrics for the outcomes or how they organize their data; hence, establishing a direct summarization of the results is not practical. We believe that future systematic reviews should also cover the clinical relevance of the methods for lung cancer applications. This review did not find any implementation of vision transformers-based methods for mobile devices. Mobile devices will carry a significant role in transforming cancer care, and porting of highly accurate and effective strategies for cancer diagnosis and classification to mobile devices will open new dimensions in future digital healthcare by facilitating ease of use and accessibility. The included studies were inconsistent in reporting the training time required for the model. For example, the reviewers could not find this information in most of the studies or did not compare how the models would behave on different hardware and whether the training/inference time would see a major reduction. It is expected that providing web-based demos for the proposed models, in general, will increase the interest of doctors, physicians, and students in exploring the potential of vision transformers for lung cancer applications. However, this review did not find web-based platforms that used vision transformers for lung cancer applications.

# 5 Strengths and Limitations
## 5.1 Strengths

With the recent popularity of vision transformer-based AI methods in medical imaging, there has been a growing interest in reviews on the topic [18], [58], [59], [60], [61]. However, we did not find any previous review on vision transformers for lung cancer imaging. This is the first review covering the classification, diagnosis, and prognosis applications of vision transformers for lung cancer imaging.

In this review, we have summarized the key vision transformer-based methods for lung cancer applications that will help the readers and the researchers to identify the potential opportunities and related challenges in developing advanced methods for lung cancer analysis. In the review, we followed the guidelines of the PRISMA-ScR [22]. We included the most relevant studies from popular scientific databases that cover technical and healthcare literature. We overcame bias in study selection by adapting an independent selection mechanism of studies by two reviewers that a third reviewer validated. We identified the key areas and gaps in the vision transformer-based methods to which researchers may contribute. To the best of our knowledge, this is the first comprehensive review that explores the role of vision transformers in improving lung cancer classification and prognosis. Furthermore, it covers the most recent studies reported by the researchers. Hence, this review will be beneficial for researchers and practitioners interested in the transformation of digital healthcare, in general, and lung cancer, in particular.

### 5.2 Limitations

Since this review covered imaging-based applications only, clinical factors and living habits of lung cancer patients were not covered in the included studies, which would otherwise provide key information in the course of the disease. We did not evaluate the code as this was beyond the scope of this review. Since the included studies varied in terms of the datasets, or the number of samples/patients used, it was impossible to establish a direct comparison of their performances on the classification or prognosis of lung cancer. This review does not provide a discussion on the training delays due to two reasons. Firstly, such information was not provided in the included studies. Secondly, different research groups may vary greatly in their access to computational resources and GPUs. We understand that the interest in using and developing newer architectures of vision transformer-based AI methods for lung cancer imaging is growing rapidly. Hence, we cannot rule out the possibility that several other studies may come out while this work is being drafted, despite our best efforts to include the most recent studies until December 2022. This review covers studies published in English, so, relevant studies in other languages (if any) are not included.

## 6 Conclusion

In this work, we undertook a scoping review of 34 studies investigating the development and implementation of AI methods in lung cancer imaging, specifically using vision transformer models. Our review work indicates that vision transformer-based methods have been developed for the classification of lung cancer types and survival prediction of lung patients. Most reported methods have achieved performance propelling forward the field of AI for lung cancer imaging. The included studies evaluated the performance in terms of accuracy, the area under the ROC curve, and the concordance index. Additionally, we cataloged publicly available datasets for lung cancer imaging. Despite these advancements, we also identified areas for improvement, such as reducing model complexity, bridging the gap between clinical practice and vision transformer-based AI methods, and increasing geographical diversity in published studies. Moreover, there is an urgent need to develop explainable vision transformer models for lung cancer imaging, as this will enhance the trust and acceptance of these methods among all stakeholders. We anticipate that our findings will provide a valuable reference text for researchers and students in the interdisciplinary fields of medical AI and cancer imaging.

**Abbreviations:**

| | |
|---|---|
| AI | Artificial Intelligence |
| AUC | Area under ROC curve |
| CNN | Convolutional Neural Networks |
| CT | Computed Tomography |
| DSC | dice similarity coefficient |
| GPU | Graphics Processing Unit |
| LIDC | Lung Imaging Database Consortium |
| MAE | mean absolute error |
| PET | Positron Emission Tomography |
| RMSE | root mean square error |
| SOTA | State-of-the-art |
| TCGA | The Cancer Genome Atlas |
| WSI | Whole slide imaging |

**Declarations**

**Ethics approval and consent to participate:** Not applicable.

**Consent for publication:** Not applicable.

**Availability of data and material:** All data generated or analysed during this study are included in this published article and its supplementary information files.

**Competing Interest:** The authors declare that they have no competing interests.

**Funding:** Not applicable.

**Authors contributions:** H. A. contributed to the conception, design, literature search, data selection, data synthesis, data extraction, and drafting. F. M. contributed to the data synthesis, data extraction, and drafting. Z. S. contributed to the drafting and critical revision of the manuscript. All authors gave their final approval and accepted accountability for all aspects of the work.

**Acknowledgements:** Not applicable.

**References**

[1] C. S. D. Cruz, L. T. Tanoue, and R. A. Matthay, "Lung cancer: epidemiology, etiology, and prevention," *Clin. Chest Med.*, vol. 32, no. 4, pp. 605–644, 2011.
[2] P. M. de Groot, C. C. Wu, B. W. Carter, and R. F. Munden, "The epidemiology of lung cancer," *Transl. Lung Cancer Res.*, vol. 7, no. 3, p. 220, 2018.
[3] S. J. Lewis, Z. Gandomkar, and P. C. Brennan, "Artificial Intelligence in medical imaging practice: looking to the future," *J. Med. Radiat. Sci.*, vol. 66, no. 4, pp. 292–295, 2019.
[4] P. Rajpurkar, E. Chen, O. Banerjee, and E. J. Topol, "AI in health and medicine," *Nat. Med.*, vol. 28, no. 1, pp. 31–38, 2022.
[5] A. Krizhevsky, I. Sutskever, and G. E. Hinton, "Imagenet classification with deep convolutional neural networks," *Commun. ACM*, vol. 60, no. 6, pp. 84–90, 2017.

[6] I. Goodfellow, Y. Bengio, and A. Courville, *Deep learning*. MIT press, 2016.
[7] A. S. Lundervold and A. Lundervold, "An overview of deep learning in medical imaging focusing on MRI," *Z. Für Med. Phys.*, vol. 29, no. 2, pp. 102–127, 2019.
[8] H. Ali *et al.*, "The role of generative adversarial networks in brain MRI: a scoping review," *Insights Imaging*, vol. 13, no. 1, pp. 1–15, 2022.
[9] T. Würfl, F. C. Ghesu, V. Christlein, and A. Maier, "Deep learning computed tomography," in *Medical Image Computing and Computer-Assisted Intervention-MICCAI 2016: 19th International Conference, Athens, Greece, October 17-21, 2016, Proceedings, Part III 19*, Springer, 2016, pp. 432–440.
[10] J. K. Min, M. S. Kwak, and J. M. Cha, "Overview of deep learning in gastrointestinal endoscopy," *Gut Liver*, vol. 13, no. 4, p. 388, 2019.
[11] P. Lakhani and B. Sundaram, "Deep learning at chest radiography: automated classification of pulmonary tuberculosis by using convolutional neural networks," *Radiology*, vol. 284, no. 2, pp. 574–582, 2017.
[12] T. Iqbal and H. Ali, "Generative adversarial network for medical images (MI-GAN)," *J. Med. Syst.*, vol. 42, pp. 1–11, 2018.
[13] A. Dosovitskiy *et al.*, "An image is worth 16x16 words: Transformers for image recognition at scale," *ArXiv Prepr. ArXiv201011929*, 2020.
[14] S. Zheng *et al.*, "Rethinking semantic segmentation from a sequence-to-sequence perspective with transformers," in *IEEE/CVF conference on computer vision and pattern recognition*, 2021, pp. 6881–6890.
[15] N. Carion, F. Massa, G. Synnaeve, N. Usunier, A. Kirillov, and S. Zagoruyko, "End-to-end object detection with transformers," in *Computer Vision–ECCV 2020: 16th European Conference, Glasgow, UK, August 23–28, 2020, Proceedings, Part I 16*, Springer, 2020, pp. 213–229.
[16] X. Gao *et al.*, "COVID-VIT: Classification of Covid-19 from 3D CT chest images based on vision transformer model," in *2022 3rd International Conference on Next Generation Computing Applications (NextComp)*, IEEE, 2022, pp. 1–4.
[17] S. Watanabe, T. Ueno, Y. Kimura, M. Mishina, and N. Sugimoto, "Generative image transformer (GIT): unsupervised continuous image generative and transformable model for [123 I] FP-CIT SPECT images," *Ann. Nucl. Med.*, vol. 35, pp. 1203–1213, 2021.
[18] K. He *et al.*, "Transformers in medical image analysis: A review," *Intell. Med.*, vol. 3, no. 1, pp. 59–78, 2022, doi: 10.1016/j.imed.2022.07.002.
[19] G. Chassagnon *et al.*, "Artificial intelligence in lung cancer: current applications and perspectives," *Jpn. J. Radiol.*, pp. 1–10, 2022.
[20] C. Kaur and U. Garg, "Artificial intelligence techniques for cancer detection in medical image processing: A review," *Mater. Today Proc.*, 2021.
[21] S. Wang *et al.*, "Artificial intelligence in lung cancer pathology image analysis," *Cancers*, vol. 11, no. 11, p. 1673, 2019.
[22] A. C. Tricco *et al.*, "PRISMA extension for scoping reviews (PRISMA-ScR): checklist and explanation," *Ann. Intern. Med.*, vol. 169, no. 7, pp. 467–473, 2018.
[23] M. Ouzzani, H. Hammady, Z. Fedorowicz, and A. Elmagarmid, "Rayyan—a web and mobile app for systematic reviews," *Syst. Rev.*, vol. 5, pp. 1–10, 2016.
[24] Y. Chen, J. Feng, J. Liu, B. Pang, D. Cao, and C. Li, "Detection and Classification of Lung Cancer Cells Using Swin Transformer," *J. Cancer Ther.*, vol. 13, no. 7, pp. 464–475, 2022.

[25] T. Aitazaz, A. Tubaishat, F. Al-Obeidat, B. Shah, T. Zia, and A. Tariq, "Transfer learning for histopathology images: an empirical study," *Neural Comput. Appl.*, 2022, doi: 10.1007/s00521-022-07516-7.

[26] Y. Zheng *et al.*, "A Graph-Transformer for Whole Slide Image Classification," *IEEE Trans. Med. Imaging*, vol. 41, no. 11, pp. 3003–3015, 2022, doi: 10.1109/TMI.2022.3176598.

[27] Z. Shao, H. Bian, Y. Chen, Y. Wang, J. Zhang, and X. Ji, "Transmil: Transformer based correlated multiple instance learning for whole slide image classification," in *Advances in neural information processing systems*, 2021, pp. 2136–2147.

[28] X. Wang *et al.*, "Transformer-based unsupervised contrastive learning for histopathological image classification," *Med. Image Anal.*, vol. 81, 2022, doi: 10.1016/j.media.2022.102559.

[29] S. Heidarian, "Capsule Network-based COVID-19 Diagnosis and Transformer-based Lung Cancer Invasiveness Prediction via Computerized Tomography (CT) Images," Doctoral thesis, Concordia University, 2022.

[30] D. Liu, F. Liu, Y. Tie, L. Qi, and F. Wang, "Res-trans networks for lung nodule classification," *Int. J. Comput. Assist. Radiol. Surg.*, vol. 17, no. 6, pp. 1059–1068, 2022, doi: 10.1007/s11548-022-02576-5.

[31] R. Wang, Y. Zhang, and J. Yang, *TransPND: A Transformer Based Pulmonary Nodule Diagnosis Method on CT Image*, vol. 13535. in Lecture Notes in Computer Science (including subseries Lecture Notes in Artificial Intelligence and Lecture Notes in Bioinformatics), vol. 13535. 2022. doi: 10.1007/978-3-031-18910-4_29.

[32] P. Wu, J. Chen, and Y. Wu, "Swin Transformer based benign and malignant pulmonary nodule classification," in *Proceedings of SPIE - The International Society for Optical Engineering*, 2022. doi: 10.1117/12.2656809.

[33] Y. Xiong, B. Du, Y. Xu, J. Deng, Y. She, and C. Chen, "Pulmonary Nodule Classification with Multi-View Convolutional Vision Transformer," in *2022 International Joint Conference on Neural Networks (IJCNN)*, 2022, pp. 1–7. doi: 10.1109/IJCNN55064.2022.9892716.

[34] J. Yang, H. Deng, X. Huang, B. Ni, and Y. Xu, "Relational Learning Between Multiple Pulmonary Nodules via Deep Set Attention Transformers," in *2020 IEEE 17th International Symposium on Biomedical Imaging (ISBI)*, 2020, pp. 1875–1878. doi: 10.1109/ISBI45749.2020.9098722.

[35] R. J. Chen *et al.*, "Scaling Vision Transformers to Gigapixel Images via Hierarchical Self-Supervised Learning," in *2022 IEEE/CVF Conference on Computer Vision and Pattern Recognition (CVPR)*, New Orleans, LA, USA: IEEE, 2022, pp. 16144–16155. doi: 10.1109/CVPR52688.2022.01567.

[36] T. Dhamija, A. Gupta, S. Gupta, Anjum, R. Katarya, and G. Singh, "Semantic segmentation in medical images through transfused convolution and transformer networks," *Appl. Intell.*, 2022.

[37] M. D. Alahmadi, "Medical Image Segmentation with Learning Semantic and Global Contextual Representation," *Diagnostics*, vol. 12, no. 7, 2022, doi: 10.3390/diagnostics12071548.

[38] D. Guo and D. Terzopoulos, "A Transformer-Based Network for Anisotropic 3D Medical Image Segmentation," in *2020 25th International Conference on Pattern Recognition (ICPR)*, 2021, pp. 8857–8861. doi: 10.1109/ICPR48806.2021.9411990.

[39] J. Jiang, N. Tyagi, K. Tringale, C. Crane, and H. Veeraraghavan, *Self-supervised 3D Anatomy Segmentation Using Self-distilled Masked Image Transformer (SMIT)*, vol. 13434. in Lecture

Notes in Computer Science (including subseries Lecture Notes in Artificial Intelligence and Lecture Notes in Bioinformatics), vol. 13434. 2022. doi: 10.1007/978-3-031-16440-8_53.

[40] S. Wang, A. Jiang, X. Li, Y. Qiu, M. Li, and F. Li, "DPBET: A dual-path lung nodules segmentation model based on boundary enhancement and hybrid transformer," *Comput. Biol. Med.*, vol. 151, p. 106330, 2022, doi: 10.1016/j.compbiomed.2022.106330.

[41] H. Xie, Z. Chen, J. Deng, J. Zhang, H. Duan, and Q. Li, "Automatic segmentation of the gross target volume in radiotherapy for lung cancer using transresSEUnet 2.5 D Network," *J. Transl. Med.*, vol. 20, no. 1, pp. 1–12, Nov. 2022.

[42] D. Yang, A. Myronenko, X. Wang, Z. Xu, H. R. Roth, and D. Xu, "T-AutoML: Automated machine learning for lesion segmentation using transformers in 3d medical imaging," presented at the Proceedings of the IEEE/CVF international conference on computer vision, 2021, pp. 3962–3974.

[43] D. -P. Dao *et al.*, "Survival Analysis based on Lung Tumor Segmentation using Global Context-aware Transformer in Multimodality," in *2022 26th International Conference on Pattern Recognition (ICPR)*, 2022, pp. 5162–5169. doi: 10.1109/ICPR56361.2022.9956406.

[44] C. Niu and G. Wang, "Unsupervised contrastive learning based transformer for lung nodule detection," *Phys. Med. Biol.*, vol. 67, no. 20, 2022, doi: 10.1088/1361-6560/ac92ba.

[45] W. Zhao *et al.*, "GMILT: A Novel Transformer Network That Can Noninvasively Predict EGFR Mutation Status," *IEEE Trans. Neural Netw. Learn. Syst.*, pp. 1–15, 2022, doi: 10.1109/TNNLS.2022.3190671.

[46] J. Shao *et al.*, "Radiogenomic System for Non-Invasive Identification of Multiple Actionable Mutations and PD-L1 Expression in Non-Small Cell Lung Cancer Based on CT Images," *Cancers*, vol. 14, no. 19, 2022, doi: 10.3390/cancers14194823.

[47] H. Wang *et al.*, "Static–Dynamic coordinated Transformer for Tumor Longitudinal Growth Prediction," *Comput. Biol. Med.*, vol. 148, 2022, doi: 10.1016/j.compbiomed.2022.105922.

[48] J. Lian *et al.*, "Early stage NSCLS patients' prognostic prediction with multi-information using transformer and graph neural network model," *eLife*, vol. 11, 2022, doi: 10.7554/eLife.80547.

[49] X. Ma, L. Xia, J. Chen, W. Wan, and W. Zhou, "Development and validation of a deep learning signature for predicting lymph node metastasis in lung adenocarcinoma: comparison with radiomics signature and clinical-semantic model," *Eur. Radiol.*, 2022, doi: 10.1007/s00330-022-09153-z.

[50] P. Song *et al.*, "MSTS-Net: malignancy evolution prediction of pulmonary nodules from longitudinal CT images via multi-task spatial-temporal self-attention network," *Int. J. Comput. Assist. Radiol. Surg.*, pp. 1–9, Nov. 2022.

[51] J. Fang *et al.*, "Siamese Encoder-based Spatial-Temporal Mixer for Growth Trend Prediction of Lung Nodules on CT Scans," in *International Conference on Medical Image Computing and Computer-Assisted Intervention*, Singapore: Springer, Sep. 2022, pp. 484–494.

[52] H. Wang, H. Zhu, and L. Ding, "Accurate Classification of Lung Nodules on CT Image Based on TransUnet," *Front. Public Health*, p. 4664, Dec. 2022.

[53] R. J. Chen *et al.*, "Multimodal Co-Attention Transformer for Survival Prediction in Gigapixel Whole Slide Images," in *Proceedings of the IEEE/CVF International Conference on Computer Vision*, 2021, pp. 3995–4005.

[54] Z. Huang, H. Chai, R. Wang, H. Wang, Y. Yang, and H. Wu, "Integration of patch features through self-supervised learning and transformer for survival analysis on whole slide

images," in *International Conference on Medical Image Computing and Computer-Assisted Intervention*, Strasbourg, France: Springer, 2021, pp. 561–570.

[55] R. Wang, Z. Huang, H. Wang, and H. Wu, "Ammasurv: asymmetrical multi-modal attention for accurate survival analysis with whole slide images and gene expression data," presented at the 2021 IEEE International Conference on Bioinformatics and Biomedicine (BIBM), IEEE, 2021, pp. 757–760.

[56] C. Li, X. Zhu, J. Yao, and J. Huang, "Hierarchical Transformer for Survival Prediction Using Multimodality Whole Slide Images and Genomics," in *26th International Conference on Pattern Recognition (ICPR)*, Montreal, QC, Canada: IEEE, 2022, pp. 4256–4262.

[57] Y. Shen *et al.*, "Explainable Survival Analysis with Convolution-Involved Vision Transformer," in *Proceedings of the AAAI Conference on Artificial Intelligence*, 2022, pp. 2207–2215.

[58] F. Shamshad *et al.*, "Transformers in medical imaging: A survey," *ArXiv Prepr. ArXiv220109873*, 2022.

[59] A. A. Akinyelu, F. Zaccagna, J. T. Grist, M. Castelli, and L. Rundo, "Brain Tumor Diagnosis Using Machine Learning, Convolutional Neural Networks, Capsule Neural Networks and Vision Transformers, Applied to MRI: A Survey," *J. Imaging*, vol. 8, no. 8, p. 205, 2022.

[60] K. Xia and J. Wang, "Recent advances of Transformers in medical image analysis: A comprehensive review," *MedComm–Future Med.*, vol. 2, no. 1, p. e38, 2023, doi: 10.1002/mef2.38.

[61] Y. Li, X. Wu, P. Yang, G. Jiang, and Y. Luo, "Machine Learning for Lung Cancer Diagnosis, Treatment, and Prognosis," *Genomics Proteomics Bioinformatics*, vol. 20, no. 5, pp. 850–866, 2022.